\documentclass[11pt,letterpaper]{article}
\pdfoutput=1
\usepackage{jheppub}
\usepackage{scalefnt}
\usepackage{xfrac}
\usepackage{caption}
\usepackage{subcaption}

\PassOptionsToPackage{jheppub}{hyperref}

\makeatletter
\everymath{\if@display\else\thickmuskip=4mu plus 4mu\fi}
\makeatother

\newcommand{\beq}{\begin{equation}}
\newcommand{\eeq}{\end{equation}}
\newcommand{\bea}{\begin{eqnarray}}
\newcommand{\eea}{\end{eqnarray}}
\newcommand{\nn}{\nonumber\\}

\def\pa{\partial}
\newcommand{\vev}[1]{\left\langle#1\right\rangle}
\newcommand{\eqn}[1]{Eq.~{\hspace{-2pt}}(\ref{#1})}
\newcommand{\eqns}[2]{Eqs.~{\hspace{-2pt}}(\ref{#1}),\,(\ref{#2})}

\newcommand{\secn}[1]{Sec.~\hspace{-2pt}\ref{#1}}

\newcommand{\reference}[1]{Ref.~\hspace{-2pt}\cite{#1}}
\newcommand{\references}[2]{Refs.~{\hspace{-2pt}}\cite{#1},\,\cite{#2}}

\newcommand{\half}{\frac{1}{2}}

\newcommand{\Deltabarbeta}{\overline{\Delta\beta}}

\newcommand{\abar}{{\overline a}}

\newcommand{\betabar}{{\overline{\beta}}}

\newcommand{\wt}[1]{\widetilde{#1}}
\def\Tr{{\rm Tr }}

\def\xip{\xi'}

\author[a,1]{Martin B Einhorn}\note{Also, \it{Michigan Center for 
Theoretical Physics, Ann Arbor, MI 48109.}} 
\author[a,b]{and D R Timothy Jones}
\affiliation[a]{Kavli Institute for Theoretical Physics,\\ University of California,
Santa Barbara, CA 93106-4030, USA}
\affiliation[b]{Dept. of Mathematical Sciences,\\ University of Liverpool, 
Liverpool L69 3BX, UK}

\emailAdd{meinhorn@umich.edu}
\emailAdd{drtj@liv.ac.uk}

\title{Grand Unified Theories in Renormalisable, Classically Scale Invariant Gravity}

\abstract{We analyze  $SO(N)$ and $SU(N)$ gauge theories with  
scalars in adjoint and fundamental representations, coupled 
to renormalisable, classically scale invariant gravity. In the specific case of $SO(12),$ 
we show that the quantum field theory can be can be asymptotically free in all 
couplings (hence ultra-violet complete). For a region of parameter space, 
Dimensional Transmutation occurs, with the adjoint vacuum expectation value breaking 
$SO(12) \to SU(6)\otimes U(1)$ and producing a Low Energy Effective Theory having 
Einstein-Hilbert gravity. We verify that certain minima are locally stable and lie 
within the catchment basin of the ultraviolet fixed points.} 


\begin{document}
\maketitle
\vspace{-10mm}
\medskip
\hrule
\vfill
\pagebreak

\section{Introduction} \label{sec:intro}

The search for a successful unification of the Standard Model (SM) with gravitational interactions 
remains a major preoccupation  for theoretical physicists. Most effort for over the last thirty years 
has been expended on string theory; with results  often of considerable mathematical elegance 
but as yet unproven physical relevance. There are other possibilities, however. In some 
ways the most conceptually simple of these (in that it preserves the renormalisability characteristic 
of the SM) is renormalisable quantum gravity {RQG}~\cite{Stelle:1976gc}, having a Lagrangian 
containing no more than two powers of the curvature. A subset of such models is 
asymptotically free (AF) in the gravitational 
interactions~\cite{Fradkin:1981hx, Fradkin:1981iu, Avramidi:1985ki}. 
(Refs.~\cite{Buchbinder:1992rb, Avramidi:2000pia} provide comprehensive reviews up to their 
publication dates.)   
If these are to include matter in a manner in which all their couplings are also AF, the content of 
such models is rather strictly circumscribed. First of all, it is well known that, in flat space, only 
non-Abelian gauge theories can be AF in their gauge couplings, and only within a certain subset of 
models having additional scalars and fermions. Further restrictions are necessary in order to have 
all the matter couplings AF as well~\cite{Cheng:1973nv}. Similarly, with renormalisable quantum gravity 
(RQG), there is only a narrowly restricted subset of such non-Abelian gauge theories that remain  
AF in all couplings~\cite{Buchbinder:1989jd, Buchbinder:1989ma}. 

An alternative approach, given the renormalisable nature of 
RQG, is that of Asymptotic Safety, (AS)~\cite{Weinberg:1979}-\cite{Percacci:2017fkn}. 
Originally suggested by Weinberg~\cite{Weinberg:1979}, the key idea of AS for a 
quantum field theory is the existence of ultra-violet fixed points (UVFP) for the 
coupling constants, rendering the theory UV complete. 
As mentioned above, it is well-known that the gravitational couplings are AF for a 
range of couplings, so it is natural to reconsider these models in the context of AS. In 
\references{Niedermaier:2009zz}{Niedermaier:2010zz}, Niedermaier introduced a new 
scheme for renormalising the Planck mass 
$M_P$ (or Newton's constant $G_N$) and the cosmological constant 
$\Lambda.$ These, like all coupling constants with positive mass dimension, are UV 
irrelevant in conventional perturbation theory. Instead, in Niedermaier's nonminimal 
renormalisation scheme, they acquire anomalous dimensions such that they  
approach finite nonzero values asymptotically.  Even more remarkably, he showed that 
these UVFPs can be calculated in perturbation theory, and, since the usual dimensionless
couplings are AF, these new UVFPs can be determined independently of those couplings. 
In \reference{Niedermaier:2010zz}, he argued that his one-loop results were unique and 
gauge-invariant.  He also showed that the same construction can also be applied to 
nonrenormalisable Einstein-Hilbert (E-H) gravity, but that the results are unavoidably 
gauge-dependent. This perturbative approach has not to our knowledge been extended 
to higher loops or to include matter fields, although both may be possible.\footnote{For 
some recent work on AS, see 
Refs.~\cite{Eichhorn:2017egq, Christiansen:2017cxa, Falls:2018ylp}, and for 
references to work up to 2017, see \reference{Percacci:2017fkn}}.

By contrast, in our work, we have followed the usual path of perturbative 
renormalisation, employing minimal subtraction, for values of the gravitational couplings 
that are AF, but for which the {\it ratios\/} of dimensionless couplings approach UVFPs. 
We consider the possibility that $M_P$ and $\Lambda$ arise dynamically as the 
couplings increase at lower energy scales, the result of a form of dimensional 
transmutation (DT) that is still calculable perturbatively. In fact, this can occur even in 
pure gravity, but unfortunately we found that the extremum was always locally unstable. 
As we shall discuss subsequently, in the past and 
at present, we have focused on whether, through interactions between gravity and 
matter, AF for all dimensionless couplings and DT may be preserved but some extrema 
become locally stable for some range of parameters. 

Attempts have been made to extend renormalisability to conformal or Weyl gravity, in 
which the only quadratic curvature term is the square of the Weyl tensor. In general, such 
models are only conformal classically, with the symmetry broken in the associated QFT 
because of the conformal anomaly. One may hope that extensions of such models that 
have local supersymmetry, for both gravity and matter, may in fact not have such an 
anomaly.~\cite{Fradkin:1982xc, Fradkin:1983tg, Fradkin:1985am}. This remains 
controversial because, in the absence of a regulator that respects all symmetries, it is not 
clear that, beyond one-loop order, conformal invariance can always be restored 
on-shell~\cite{Fradkin:1983tg, Fradkin:1985am}.\footnote{In fact, there is evidence to the 
contrary at two-loop order~\cite{Fradkin:1985am}. Note that with supergravity, 
most authors do not include quadratic terms but deal only with the effective field theory 
having the locally supersymmetric counterpart of the Einstein-Hilbert action plus matter. 
In this respect, the two references 
\cite{Fradkin:1983tg, Fradkin:1985am}\ are exceptional.}

Our interests are not in a conformal version of these models having zero beta-functions. 
In the absence of anything but quadratic curvature terms, RQG is classically scale invariant, 
but the associated QFT does have a conformal anomaly that breaks this global symmetry. 
This suggests that an interesting subset of models may be those having 
no mass parameters in the action, i.e., possessing only terms that are classically 
scale-invariant~\cite{Salvio:2014soa, Einhorn:2014gfa}. Masses will be generated by DT, 
either nonperturbatively as in Yang-Mills theory or perturbatively in a manner somewhat 
analogous to massless electrodynamics~\cite{Coleman:1973jx}. We are especially interested 
in models that have all couplings 
AF~\cite{Giudice:2014tma, Einhorn:2015lzy, Einhorn:2016mws} in perturbation theory.

In a  previous paper~\cite{Einhorn:2016mws}, we analysed an $SO(10)$ gauge theory 
coupled to RQG, including a scalar multiplet in the adjoint representation. We showed, that if 
we impose scale invariance on the classical theory, it can undergo DT and that there 
is a region of parameter space such that the couplings are all AF\footnote{To obtain 
asymptotic freedom of all the matter couplings it is necessary for the one-loop contribution to 
the gauge $\beta$-function be (nearly) as small in magnitude as possible, as well as, 
obviously, having the correct sign. This we assumed results from an appropriate contribution 
to it from fermions.}, remaining perturbatively small at scales down to the point where DT 
occurs. At this scale the vacuum expectation value (vev) of the scalar multiplet breaks 
$SO(10)$ to $SU(5)\otimes U(1)$, and the effective theory at lower scales has an Einstein 
$R$ term arising from the dimensionless nonminimal coupling of scalars to gravity. 

Elsewhere~\cite{Einhorn:2017jbs}, we  revisited the classic flat space $\beta$-function
calculations of CEL~\cite{Cheng:1973nv}, which were motivated by the search for
asymptotically free (and hence UV  complete) non-Abelian gauge theories.   We
found a number of errors in the CEL results, including the cases  of
both a $SO(N)$ and a $SU(N)$ gauge theory coupled to two scalar
multiplets: an adjoint and a fundamental. Here we extend those results
to  classically scale invariant RQG to see whether the UV completeness and DT 
phenomena we identified in~\cite{Einhorn:2016mws}\ persist for these cases, 
which we regard as prototypes for a realistic Grand Unified Theory (GUT) due to 
the more complicated scalar sector. We consider in detail both the case $SO(12)$ and the 
limit of large $N$ for both $SO(N)$ and $SU(N)$.  

\section{The $SO(N)$ theory.}

The action  we shall consider is 
\beq\label{eq:sho}
S_{tot} = \int d^4x\sqrt{g} \left[ -\frac{1}{4}(G^a_{\mu\nu})^2
-\half g^{\mu\nu}D_{\mu}\phi^a D_{\nu}\phi^a 
-\half g^{\mu\nu} D_{\mu}\chi^i D_{\nu}\chi^i-V(\phi,\chi) \right] + S_{grav}
\eeq
where 
\beq\label{eq:sgrav}
S_{grav} = \int d^4x\sqrt{g} \left[\frac{C^2}{2a}+\frac{R^2}{3b}+cG. 
\right]
\eeq
Here $C$ is the Weyl tensor and $G$ the Gauss-Bonnet term; $a$, $b$, $c$
are dimensionless coupling constants. 

All the fields are real. $\chi^i$ is in the defining (fundamental) representation 
with $i:1 \cdots N$ and $\phi^a$ is in the adjoint representation 
with $a: 1\cdots N(N-1)/2$. 
It is convenient for some purposes to write $\phi^a$ 
as $\Phi = T^a\phi^a$ where $T^a$ are the $N\times N$ 
hermitian generators of $SO(N)$, normalised so that 
\beq\label{TRnorm}
\Tr \left[T^a T^b\right]=\half\delta^{ab}.
\eeq
For an arbitrary irreducible representation of the generators $R^a$, we write 
\beq
\Tr \left[R^a R^b\right]=T_R\delta^{ab}.
\eeq
$T_R$ is determined for each representation by the normalisation \eqn{TRnorm}.

The general form of the scale invariant scalar potential
is\footnote{CEL's adjoint has $\phi\equiv \phi_a B^a,$ with
normalization $\Tr[\phi^2]{=}-\phi_a\phi_a/2{=}-T_2/2.$  ($B_a=iR_a,$ 
$\phi=i\Phi/\sqrt2.$)} 
\beq\label{eq:classpot6}
V(\Phi,\chi)=\frac{\lambda_1}{2}T_2^2
+\lambda_2 T_4+\frac{\lambda_3}{8}(\chi^t\chi)^2 
+\frac{\lambda_4}{2}T_2(\chi^t\chi)+
\frac{\lambda_5}{4}\chi^t\Phi^2\chi 
- \xi_1T_2 R-\half \xi_2\chi^t\chi R,
\eeq
where $T_m=\Tr[\Phi^m]=\Tr[(\phi_a T^a)^m].$ In particular,
$T_2=\Tr[\left(\phi_a T^a\right)^2]=\frac{1}{2}\sum_a\left(\phi_a\right)^2.$ 

\section{The $SO(N)$  $\beta$-functions}

\subsection{Classical gravity} 

Although quantum corrections due to gravity are not included in this approximation, the 
$\xi_{1,2}$ terms in the potential, \eqn{eq:classpot6}, must still be present since,
even in a background gravitational field having $R\neq0,$ $\xi_{1,2}$ undergo renormalisation. 

The flat space $\beta$-functions are given in \eqn{eq:betaflat}. (We suppress throughout 
a factor of $(16\pi^2)^{-1}$ in each $\beta$-function.) 
\begin{subequations}\label{eq:betaflat}
\begin{align}
\beta_{g^2}&=-b_g (g^2)^2,\ b_g\equiv 
2\left(\frac{11}{3}C_G-\frac{1}{6}T_S -\frac{2}{3}T_F\right),
\label{eq:betaalpha1}\\
\begin{split}
\beta_{\lambda_1}^{(0)}&=\frac{(N(N{-}1)+16)}{2}\lambda_1^2+
 2(2N{-}1)\lambda_1\lambda_2+6\lambda_2^2+N\lambda_4^2 + \lambda_4\lambda_5\\
 &\hskip0.5in  +9g^4{-}6(N{-}2)g^2\lambda_1,  
 \end{split}\\
 \begin{split}
 \beta_{\lambda_2}^{(0)}&= 
(2N{-}1)\lambda_2^2{+}12\lambda_1\lambda_2{+}\frac{\lambda_5^2}{8}
+\frac{3(N{-}8)g^4}{2}{-}6(N{-}2)g^2\lambda_2,
\end{split}\\
\beta_{\lambda_3}^{(0)}&=(N{+}8)\lambda_3^2{+}
(N{-}1)\Big(\frac{N\lambda_4^2}{2}+
\frac{\lambda_4\lambda_5}{2}+\frac{\lambda_5^2}{16}\Big)
+\frac{3(N{-}1)g^4}{4}{-}3(N{-}1)g^2\lambda_3,\\
\begin{split}
\beta_{\lambda_4}^{(0)}&=4\lambda_4^2+\lambda_4\bigg[
+\frac{(N(N{-}1){+}4)}{2}\lambda_1+(2N{-}1)\lambda_2+(N{+}2)\lambda_3\bigg]\\
&\hskip5mm +\lambda_5\bigg[ \frac{(N{-}1)}{4}\lambda_1+
\frac{1}{2}\lambda_2+\frac{1}{2}\lambda_3\bigg]+\frac{1}{8}\lambda_5^2
+\frac{3g^4}{2}-\frac{3(3N{-}5)g^2\lambda_4}{2},
\end{split}\\
\beta_{\lambda_5}^{(0)}&=\frac{N}{4}\lambda_5^2 +
\lambda_5\bigg[2\lambda_1+(N{-}1)\lambda_2+2\lambda_3+
8\lambda_4\bigg]+3(N{-}4)g^4-\frac{3(3N{-}5)g^2\lambda_5}{2}.
\end{align}
\end{subequations}
In \eqn{eq:betaalpha1}, $T_{S,F}$ represent the values of $T_R$ for the scalars and fermions.  For 
$SO(N),$ $C_G=(N{-}2)/2.$  For this model, $T_S=(N{-}2)/2+1/2=(N{-}1)/2,$ so
 \beq
 b_g=\frac{21N-43}{6}-\frac{4}{3}T_F.
 \eeq
 
Although $\{\xi_1,\xi_2\}$  are not relevant for flat space, in curved spacetime, matter 
self-interac\-tions contribute to their one-loop $\beta$ functions. Letting 
$\xi'_i\equiv\xi_i + {1}/{6},$ we have
\begin{subequations}\label{eq:betaxip}
\begin{align}
\beta_{\xi'_1}^{(0)}&=\beta_{11}\xi'_1+\beta_{12}\xi'_2,\\
\beta_{\xi'_2}^{(0)}&=\beta_{21}\xi'_1+\beta_{22}\xi'_2,
\end{align}
\end{subequations}
where
\begin{subequations}\label{eq:betaij2}
\begin{align}
\beta_{11}&=\frac{N(N{-}1){+}4}{2}\,\lambda_1{+}
(2N{-}1)\lambda_2-3(N{-}2)\alpha,\\
\beta_{12}&=N\lambda_4{+}\frac{1}{2}\lambda_5,\\
\beta_{21}&=\frac{(N{-}1)}{2}\left[N\lambda_4+\frac{1}{2}\lambda_5\right],\\
\beta_{22}&=(N{+}2)\lambda_3-\frac{3(N{-}1)}{2}\alpha.
\end{align}
\end{subequations}
Here (and subsequently) we have set $\alpha \equiv g^2$.  \eqn{eq:betaxip} obviously gives rise to    
fixed point solutions for $\beta_{\xi'_{1,2}}^{(0)}$ corresponding to conformal coupling, 
$\xi'_1{=}\xi'_2{=}0.$ Assuming that $\det[\beta_{ij}]\neq 0,$ there are no other fixed point solutions.  
We shall see shortly that the one-loop gravity contribution changes that result.

\subsection{Gravitational Corrections}

The gravitational corrections to the matter $\beta$-functions have a
universal form.  We infer\footnote{A word of caution concerning 
\reference{Salvio:2014soa}: At the time of our writing, the journal version differs from the 
earlier arXiv version, correcting some formulas.} from \reference{Salvio:2014soa} that 
the full $\beta$-functions are
\begin{subequations}\label{eq:betamatter}
\begin{align}
\beta_{\lambda_1}&=\beta_{\lambda_1}^{(0)}
+\Delta\beta^{(1)}(\xi'_1)+\lambda_1\Delta\beta^{(2)}(\xi'_1),\\
\beta_{\lambda_2}&=\beta_{\lambda_2}^{(0)}
+\lambda_2\Delta\beta^{(2)}(\xi'_1),\\
\beta_{\lambda_3}&=\beta_{\lambda_3}^{(0)}
+\Delta\beta^{(1)}(\xi'_2)+\lambda_3\Delta\beta^{(2)}(\xi'_2),\\
\beta_{\lambda_4}&=\beta_{\lambda_4}^{(0)}
+\Delta\beta^{(4)}(\xi'_1,\xi'_2)+\lambda_4\Delta\beta^{(3)}(\xi'_1,\xi'_2),\\
\beta_{\lambda_5}&=\beta_{\lambda_5}^{(0)}
+\lambda_5\Delta\beta^{(3)}(\xi'_1,\xi'_2),\\
\beta_{\xi'_j}&=\beta_{\xi'_j}^{(0)}+\Delta\beta_{\xi'_j}(\xi'_j),
\end{align}
\end{subequations} 
where 
\begin{subequations}\label{eq:betamatterA}
\begin{align}
\Delta\beta^{(1)}(\xi')&=\!a^2\Big(\xi'{-}\frac{1}{6}\Big)^{\!2}
\left[5\!+\! 9\xi'^{\,2} x^2 \right] ,\\
\Delta\beta^{(2)}(\xi')&=a\left(5{-}18x\xi'^{\,2}\right),\\
\Delta\beta^{(3)}(\xi'_1,\xi'_2)&=a\left[5-3x\!\left(\xi'_1{}^2{+}\xi'_2{}^2{+}4\,\xi'_1\xi'_2
\right)\right],\!\\
\Delta\beta^{(4)}(\xi'_1,\xi'_2)&\!=
a^2\Big(\xi'_1{-}\frac{1}{6}\Big)\Big(\xi'_2{-}\frac{1}{6}\Big)
\left[ 5{+}9x^2\xi'_1\xi'_2\right],\\
\label{deltabetaxip}
\Delta\beta_{\xi'}(\xi')&\!=a\Big(\xi'{-}\frac{1}{6}\Big)\!\! 
\left(\frac{10}{3 x}-\frac{3}{2} \xi'(2\xi'{+}1) x\!\right).
\end{align}
\end{subequations} 
The $\beta$-function for the gauge coupling is unchanged by gravity (at one loop) 
and remains as in \eqn{eq:betaalpha1}.

For the gravitational self-interactions  we have (again of universal form):
\begin{subequations}
\begin{align}
\label{eq:betaabc}
\beta_a&=-b_2 a^2,& \beta_b&=-a^2 b_3(x,\xi'_1,\xi'_2),& \beta_c&{=}-b_1,\!\\
\label{eq:b1b2b3}
b_2&= \frac{133}{10}+ N_a,&
b_3&\!\equiv\! \frac{10}{3}{-}5 x{+}\frac{5x^2}{12}{+}
\left(\frac{N(N{-}1)}{2}\xi'_1{}^2{+}
N\xi'_2{}^2\right)\!\frac{3x^2}{2},&
b_1&{=}\frac{196}{45}+ N_c,
\end{align}
\end{subequations} 
where $N_a{=}\left[ N_0{+}6N_{1/2} {+}12N_1^{0}{+}13N_1\right]\!/60;$
$N_c{=}\left[N_0{+}11N_{1/2}{+}62N_1^{0}{+}63N_1 \right]\!/360.$
Here $N_0$ denotes the number of (real) scalars; $N_{1/2}$, Dirac 
fermions;  $N_1^0$, massless vectors; $N_1$, massive vectors. (For
chiral or Majorana fermions,  the coefficients of $N_{1/2}$ would be
half those given above). In the last equation, we set $x\equiv b/a=1/w.$
It is sometimes convenient to use $x$ instead of $b$, with
\beq
\beta_x = a(b_2 x- b_3).
\eeq
We shall eventually be interested in determining whether this model is
asymptotically free in all couplings, requiring reduced couplings to
approach UVFP's. In this case, we have $N_0=N(N{-}1)/2{+}N=N(N{+}1)/2,$
$N_1^0=N(N{-}1)/2,$ so $N_a\geq (13N(N{-}1)/2{+}N)/60.$ For $N\geq10,$
$N_a\geq 9.92.$ So $b_2\gtrsim 23.22.$

To analyse the $\beta$-functions and seek FPs, it is convenient  to introduce 
rescaled couplings $\abar\equiv a/\alpha, x_i \equiv \lambda_i/\alpha,$ 
with consequent  ``reduced'' $\beta$-functions,
\beq\label{eq:reduced}
\betabar_\abar \equiv  \frac{d\abar}{du} = \abar(b_g-b_2\abar),\ 
\betabar_x \equiv \frac{dx}{du}= \abar(b_2 x -b_3),\  
\betabar_{x_i}\equiv\frac{dx_i}{du},\ \  \hbox{where}\ du\equiv \alpha(t)dt.
\eeq
For $b_g > 0$, as is required for asymptotic freedom of $\alpha$, we see 
that $\abar$ has a UVFP at 
\beq\label{abarfp}
\abar_{FP} = b_g/b_2.
\eeq
Hence $a\to 0$  at high energies, and, since UVFPs only exist when $b_g$ is {\it small}, 
$\abar_{FP}$ is small.   

The $\betabar_{x_i}$ are given by 
\begin{subequations}\label{redSON}
\begin{align}
\betabar_{x_1} &{=} \Big(\frac{N(N{-}1)+16}{2}\Big)x_1^2
+6x_2^2+2(2N{-}1)x_1 x_2 {+}Nx_4^2+x_4x_5 \nn
&{+}\big(b_g{-}6(N{-}2)\big)x_1 {+} 9\!
+\Deltabarbeta^{(1)}(\xi'_1)+x_1\Deltabarbeta^{(2)}(\xi'_1),\\
\betabar_{x_2} &= (2N{-}1)x_2^2 +12x_1x_2+ \frac{1}{8} x_5^2
+ \big(b_g{-}6(N{-}2)\big)x_2\nn
&+\frac{3(N{-}8)}{2}+x_2\Deltabarbeta^{(2)}(\xi'_1),\\ 
\betabar_{x_3} &= (N{+}8)x_3^2{+}\frac{N(N{-}1)}{2}x_4^2
{+}\frac{N{-}1}{16}x_5^2{+}\frac{N{-}1}{2}x_4x_5\nn
&{+}\big(b_g {-}3(N{-}1)\big)x_3 {+}
\frac{3(N{-}1)}{4}\! +\Deltabarbeta^{(1)}(\xi'_2)+x_3\Deltabarbeta^{(2)}(\xi'_2),\\
\betabar_{x_4} &= 4x_4^2+\frac{1}{8}x_5^2
+x_5\bigg[\frac{N{-}1}{4}x_1{+}\half x_2 {+}\half x_3\bigg]\!+ 
x_4\bigg[\Big(\frac{N(N{-}1)}{2}{+}2\Big)x_1{+}(2N{-}1)x_2\nn 
&+ (N{+}2)x_3 + b_g-\frac{3(3N{-}5)}{2}\bigg] +\frac{3}{2} 
 +\Deltabarbeta^{(4)}(\xi'_1,\xi'_2)+x_4\Deltabarbeta^{(3)}(\xi'_1,\xi'_2),\\
\betabar_{x_5} &= \frac{N}{4}x_5^2+x_5\bigg[2x_1{+}(N{-}1)x_2
+2x_3 +8x_4 +b_g-\frac{3(3N{-}5)}{2}\bigg]\nn
&+3(N{-}4)
+x_5\Deltabarbeta^{(3)}(\xi'_1,\xi'_2).
\end{align}
\end{subequations}
In \eqn{redSON}, the $\Deltabarbeta$ terms are identical to the $\Delta\beta$ 
terms in \eqn{eq:betamatterA}, except that $a$ is replaced by $\abar.$ 

In terms of reduced couplings, the classical gravity
$\beta$-functions for $\beta_{\xi'_{1,2}}$ become
\begin{subequations}\label{eq:betaxipred}
\begin{align}
\betabar_{\xi'_1}^{(0)}&=\betabar_{11}\xi'_1+\betabar_{12}\xi'_2,\\
\betabar_{\xi'_2}^{(0)}&=\betabar_{21}\xi'_1+\betabar_{22}\xi'_2,
\end{align}
\end{subequations}
where
\begin{subequations}\label{eq:betaij2A}
\begin{align}
\betabar_{11}&=\frac{N(N{-}1){+}4}{2}\,x_1{+}
(2N{-}1)x_2-3(N{-}2),\\
\betabar_{12}&=Nx_4{+}\frac{1}{2}x_5,\\
\betabar_{21}&=\frac{(N{-}1)}{2}\left[Nx_4+\frac{1}{2}x_5\right],\\
\betabar_{22}&=(N{+}2)x_3-\frac{3(N{-}1)}{2}.
\end{align}\end{subequations}
Adding gravitational corrections, we have
\beq\label{betabarxipk}
\betabar_{\xi'_j}=\betabar_{\xi'_j}^{(0)}+\Deltabarbeta_{\xi'_j}(\xi'_j),
\eeq
where $\Deltabarbeta_{\xi'}(\xi')$ is given by \eqn{deltabetaxip} with $a\to\abar:$
\beq\label{deltabetabarxipk}
\Deltabarbeta_{\xi'}(\xi')\! =\abar\Big(\xi'{-}\frac{1}{6}\Big)\!\! 
\left(\frac{10}{3 x}-\frac{3}{2} \xi'(2\xi'{+}1) x\!\right).
\eeq
While the matter contributions $\betabar_{\xi'_j}^{(0)}$ vanish for conformal coupling 
$(\xi'=0$ or $\xi=-1/6),$ the gravitational corrections vanish for minimal coupling, 
$(\xi'=+1/6$ or $\xi=0).$ When the two are combined, the FP respects neither limit.

\section{The $SO(12)$ Fixed Points}\label{sec:so12}

Our first goal in this class of theories is the identification of cases
such that the renormalisation group evolution of the couplings
approaches a fixed point at high energies (UVFP). In the flat space
limit, we showed that a minimum value of N=12 is required. For $N=12$ 
and with the minimum possible value of $b_g$, which is $b_g=1/6$, 
we found a UVFP with results for the quartic couplings in the flat space limit as 
follows:
\bea
x_1 &=& 0.262953,\
x_2 = 0.111668,\
x_3 = 0.376914,\nn
x_4 &=& 0.104270,\
x_5 = 0.581883.
\eea
As mentioned earlier, even with classical gravity, $\xi'_{1,2}$ both nevertheless undergo renormalisation. 
However, it is easy to see from \eqn{eq:betaxip} that they have a FP for the conformal values
\beq
\xi'_{1} = \xi'_{2} = 0.
\eeq

For curved space we expect to find a FP with similar values of the
quartic couplings and values of $\xi'_{1,2}$ both close to the conformal
value (zero). This is basically because, as explained in 
\reference{Einhorn:2016mws}, the FP result for $a/\alpha$ is necessarily small.
The result for $N=12$ is
\bea\label{uvfpgrav}
x_1 &=& 0.263283,\
x_2 = 0.111708,\
x_3 = 0.377518,\nn
x_4 &=& .104565,\
x_5 = 0.582159\nn
\xi'_1 &=& -1.41379 \times 10^{-6}, \ \xi'_2 = -2.00257\times  10^{-6}\\
x &=& 153.548\nn
\abar &=& 1/354.\nonumber
\eea
The precise values of $\abar, x$ depend on the nature of the fermion content.
To achieve $b_g = 1/6,$ we have taken $52$ two-component fermions in the
fundamental representation, so that $N_{\frac{1}{2}}= 26N = 312$.\footnote{Even 
for fixed $b_g$, there are several possible choices for the fermion representations, so 
the precise value of $N_{\frac{1}{2}}$ can vary somewhat. It is clear, however, from 
\eqns{eq:betaabc}{abarfp}\ that any such choice results in a relatively large value for $b_2$,  
and consequently a small value for $\abar.$  Thus, the results \eqn{uvfpgrav}\ will be rather  
insensitive to such variations in $N_{\frac{1}{2}}$.} Then with $N_1^0 = 66$ and $N_0 = 78,$ 
we have $b_2 = 59$, and $\abar = b_g/b_2 = 1/354$.

Note that the large value of $x$ at the FP does not invalidate perturbation theory since $b=ax$ is 
only $O(1)$ at the FP. 

There is another FP with similar values of $x_i, \xi'_{1,2}$ and a small
value of $x$, which is a  saddle point. We shall not elaborate on that.

\section{The $SO(N)$ large $N$ limit}

As we have seen, requiring the existence of a  UVFP leads to a minimum
value of $N$. It is therefore natural to  consider the large $N$
limit, first discussed in non-abelian gauge theories by 't Hooft. To
retain finite couplings in the  large $N$ limit, we need to rescale the
couplings. Since we cannot solve the theory exactly in this limit,
we shall have to require that these rescaled couplings remain 
perturbatively small at all relevant scales.   

Assuming $b_g=\wt{b}_g N,$ we found that for the quartic scalar couplings,  
scaling behavior requires
\beq\label{eq:rescalingflat}
\ \alpha=\frac{\wt{\alpha}}{N},\ 
\lambda_1=\frac{\wt{\lambda}_1}{N^2},\  
\lambda_2=\frac{\wt{\lambda}_2}{N},\ 
\lambda_3=\frac{\wt{\lambda}_3}{N},\  
\lambda_4=\frac{\wt{\lambda}_4}{N^{p_4}},\ 
\lambda_5=\frac{\wt{\lambda}_5}{N},
\eeq
for $3/2\leq p_4\leq 2.$  In \reference{Einhorn:2017icw}, we argued that the ambiguity in 
the rescaling of $\lambda_4$ reflects a nonuniformity of the limiting behavior that
is best resolved by setting $p_4=2.$

\subsection{Gravitational coupling rescaling-A}

For the remaining couplings associated with the gravitational interactions, the 
natural rescaling takes the form 
\beq\label{eq:rescalinggravA}
\ a=\frac{\wt{a}}{N}\alpha,\ 
b=\frac{\wt{b}}{N}\alpha,\  
c=N\wt{c},\ 
\xi'_1=\wt{\xi'_1},\  
\xi'_2=\wt{\xi'_2},
\eeq
We have incorporated both the rescaling with powers of $N$ and forming ratios of couplings to 
$\wt{\alpha}$ in the above equations. Note that since the scalings of $a$ and $b$ are identical, 
we have $\wt{x}=x$. 

Defining $\wt{y}_n\equiv \wt{\lambda}_n/\alpha,$ the reduced $\beta$-functions in the large $N$ 
limit are as follows: For the quartic couplings, 
\begin{subequations}\label{eq:betabarflatNA}
\begin{align}
\betabar_{\wt{y}_1}&=\frac{1}{2}\wt{y}_1^2+6\wt{y}_2^2
+9+\big(4\wt{y}_2+\wt{b}_g-6\big)\wt{y}_1,\label{sony1}\\
 \betabar_{\wt{y}_2}&= 
2\wt{y}_2^2+\frac{3}{2}+(\wt{b}_g-6)\wt{y}_2,\label{sony2}\\
\betabar_{\wt{y}_3}&=\wt{y}_3^2{+}
\frac{1}{16}\wt{y}_5^2
+\frac{3}{4}+(\wt{b}_g{-}3)\wt{y}_3,\label{sony3}\\
\betabar_{\wt{y}_4}&= \wt{y}_5\bigg[ \frac{1}{4}\wt{y}_1+
\frac{1}{2}\wt{y}_2+\frac{1}{2}\wt{y}_3\bigg]
+\frac{1}{8}\wt{y}^2_5+\frac{3}{2}+
\bigg[\frac{1}{2}\wt{y}_1+ 2\wt{y}_2+
\wt{y}_3+\wt{b}_g-\frac{9}{2}\bigg]\wt{y}_4,\label{sony4}\\
\betabar_{\wt{y}_5}&=\frac{1}{4}\wt{y}_5^2+3+
\Big(\wt{y}_2+\wt{b}_g-\frac{9}{2}\Big)\wt{y}_5,\label{sony5}
\end{align}
\end{subequations}
and for the gravitational couplings,
\begin{subequations}\label{eq:betabarcurvedNA}
\begin{align}
\betabar_{\wt{a}} &= \wt{a}(\wt{b}_g -\wt{b}_2\wt{a}),\\
\betabar_{\wt{x}} &=  -\wt{a}\,\wt{x} \Big(\frac{3}{4}\wt{\xi'_1}^2 \wt{x}
-\wt{b}_2\Big)\label{xNtilde},\\
\betabar_{\wt{\xi'_1}} &=\Big(\frac{1}{2}\wt{y}_1 +2\wt{y}_2-3\Big)\wt{\xi'_1},\label{sonxi1}\\
\betabar_{\wt{\xi'_2}} &=\Big(\wt{y}_3-\frac{3}{2}\Big)\wt{\xi'_2}
+\Big(\frac{1}{2}\wt{y}_4+\frac{1}{4}\wt{y}_5\Big)\wt{\xi'_1}.
\end{align}
\end{subequations}
Here we have defined $\wt{b}_g = b_g/N$ and 
$\wt{b}_2 = b_2/N$.

After the rescaling chosen in \eqns{eq:rescalingflat}{eq:rescalinggravA}, the
$\beta$-functions  for the quartic couplings are the same as in flat
space; hence their values at any FP will be the same as in flat space. 
 
We found UVFPs for a range of values of $\wt{b}_g$ from \eqn{eq:betabarflatNA} above in
\reference{Einhorn:2017jbs} with results as shown in
Table~\ref{soUVFPs}. (There was an calculational error in
\reference{Einhorn:2017jbs} so that  the results in Table~\ref{soUVFPs}
differ slightly from the corresponding  Table there).
It is easy to see that these remain FPs  if 
the remaining  couplings  are  $\wt{\xi'_1} =
\wt{\xi'_2} = \wt{x} = 0$. However, from \eqn{xNtilde}\ we then see  that
these  FPs are  destabilised by the gravitational corrections since  
$\wt{b}_2 > 13/120.$

\begin{table}[hbt]
\begin{center}
\begin{tabular}{|c|c|c|c|c|c|} 
\hline & & & & & \\
 $\wt{b}_g$ &
$\wt{y}_1$  & $\wt{y}_2$ & $\wt{y}_3$ & $\wt{y}_4$ & $\wt{y}_5$ \\
\hline 
0.& 2.64270& 0.27527 & 0.28941  & 0.97035 &0.74275\\ \hline 
1/6 & 2.94605& 0.28499 & 0.31255  & 1.20422 &0.77847\\ \hline
$1/3$& 3.45350 & 0.29553 & 0.34039 & 1.69063 &0.81820\\ \hline 
5/12 & 4.08657 & 0.30114 & 0.35661  & 2.51800 &0.83981\\
 \hline \hline
\end{tabular} 
\caption{\label{soUVFPs} Flat space UVFPs for $SO(\infty).$}
\end{center} 
\end{table} 

Note, however, that the results in Table~\ref{soUVFPs} correspond to
various {\it specific\/} values of  $\wt{b}_g$. There are a finite
possible number of values of $b_g$ for any given $N$, but the
number of values of $\wt{b}_g$ tends to infinity in the large $N$ limit.
 We may therefore allow arbitrary values of $\wt{b}_g$ in the search for
a FP. 

We now seek a FP with nonzero $\wt{\xi'_1}$. We may first solve 
\eqn{sony1}, \eqn{sony2} and \eqn{sonxi1} for $\wt{y}_1$, $\wt{y}_2$ and $\wt{b}_g$. 
We can then solve Eqs.~(\ref{sony3})-(\ref{sony5}) for $\wt{y}_3, \wt{y}_4, \wt{y}_5$.
This leads to two new FPs, shown in  Table~\ref{newsoUVFPs}. Unfortunately, neither of 
these FPs is UV stable. 

\begin{table}[hbt]
\begin{center}
\begin{tabular}{|c|c|c|c|c|c|c|c|} 
\hline 
& & & & & & & \\
$\wt{b}_g$ & $\wt{y}_1$  & $\wt{y}_2$ & $\wt{y}_3$ & $\wt{y}_4$ 
& $\wt{y}_5$ & $x$ & $\wt{\xi'_2}$\\
\hline 
& & & & & & &  \\
0.40961 & 4.79737 &  0.30066 &  0.35517  & 3.90016 & 0.837923 &
${\wt{b}_2}/{(\wt{\xi'_1})^2}$ & $1.88636\wt{\xi'_1}$\\ 
\hline 
& & & & & & &  \\
0.40961 & 4.79737 &  0.30066& 2.23523&    -3.19276 & 0.837923 &
${\wt{b}_2}/{(\wt{\xi'_1})^2}$ & $1.88636\wt{\xi'_1}$\\ 
\hline 
\end{tabular} 
\caption{\label{newsoUVFPs} Curved Space UVFPs for $SO(\infty).$}
\end{center} 
\end{table} 

We also have the possibility $\wt{\xi'_1} = x =0$, $\wt{\xi'_2}\neq 0$, if $\wt{y}_3 =3/2$. 
In this case however, we find no real FPs.

So it seems that with the most natural  rescaling (in terms of powers of $N$), 
gravitational corrections destabilise the UVFP. In the next subsection  
we explore an alternative rescaling, albeit with a similar conclusion.

\subsection{Gravitational coupling rescaling-B}\label{subsec:NB}

There is an alternative rescaling giving rise to a non-trivial large $N$
limit as follows:
\beq\label{eq:rescalinggravB}
\ a=\frac{\wt{a}}{N}\alpha,\ 
b=\frac{\wt{b}}{N^3}\alpha,\  
c=N\wt{c},\ 
\xip_1=N\wt{\xi'_1},\  
\xip_2=N\wt{\xi'_2},
\eeq
The quartic coupling rescaling is done in the same way as in the previous subsection.
Note that now $\wt{x}= N^2 x$. 
The resulting reduced $\beta$-functions (in the large $N$ limit) take the form
(where we have $\wt{y}_n\equiv \lambda_n/\alpha$ as before),
\begin{subequations}\label{eq:betabarflatNB}
\begin{align}
\betabar_{\wt{y}_1}&=\frac{1}{2}\wt{y}_1^2+6\wt{y}_2^2
+9+\big(4\wt{y}_2+\wt{b}_g-6\big)\wt{y}_1 
+5\wt{a}^2\wt{\xi'_1}^2 ,\\
\betabar_{\wt{y}_2}&= 
2\wt{y}_2^2+\frac{3}{2}+(\wt{b}_g-6)\wt{y}_2,\\
\betabar_{\wt{y}_3}&=\wt{y}_3^2{+}
\frac{1}{16}\wt{y}_5^2
+\frac{3}{4}+(\wt{b}_g{-}3)\wt{y}_3,\\
\betabar_{\wt{y}_4}&= \wt{y}_5\bigg[ \frac{1}{4}\wt{y}_1+
\frac{1}{2}\wt{y}_2+\frac{1}{2}\wt{y}_3\bigg]
+\frac{1}{8}\wt{y}^2_5+\frac{3}{2}\nn
&+\bigg[\frac{1}{2}\wt{y}_1+ 2\wt{y}_2+
\wt{y}_3+\wt{b}_g-\frac{9}{2}\bigg]\wt{y}_4, 
+5\wt{a}^2\wt{\xi'_1}\wt{\xi'_2}, \\
\betabar_{\wt{y}_5}&=\frac{1}{4}\wt{y}_5^2+3+
\Big(\wt{y}_2+\wt{b}_g-\frac{9}{2}\Big)\wt{y}_5,\\
\betabar_{\wt{\xi'_1}} &= \frac{1}{36}(-108+18\wt{y}_1
+72\wt{y}_2)\wt{\xi'_1}
+ \frac{10}{3}\frac{\wt{a}\wt{\xi'_1}}{\wt{x}},\\
\betabar_{\wt{\xi'_2}} &= \frac{1}{36}((-54+36\wt{y}_3)\wt{\xi'_2}
+18\wt{\xi'_1}(\wt{y}_4+\frac{1}{2}\wt{y}_5))
+ \frac{10}{3}\frac{\wt{a}\wt{\xi'_2}}{\wt{x}},\\
\betabar_{\wt{x}} &=  -(1/12)\wt{a} (9\wt{x}^2\wt{\xi'_1}^2 
-12\wt{b}_2\wt{x}+40).
\end{align}
\end{subequations}
Unlike the previous case, there is no FP  corresponding to
Table~\ref{soUVFPs} with  $\xip_1=\xip_2=x=0$. However, with
$\xip_1=\xip_2 = 0,$ we do reproduce  Table~\ref{soUVFPs}\ with a FP at
$\wt{x} = 10/(3\wt{b}_2)$. However, once again this is  not a UVFP.

\section{The $SU(N)$ theory}

In this case we have the scalar potential
\bea\label{sunpot}
V(\Phi,\chi) &= \frac{1}{2}\lambda_1 (\Tr\, \Phi^2)^2
+ \lambda_2\Tr\, \Phi^4
+\frac{1}{2}\lambda_3 (\chi_i^{\dagger}\chi^i)^2\nn
&+ \lambda_4\chi_i^{\dagger}\chi^i\Tr\, \Phi^2
+\lambda_5\chi^{\dagger}_i \Phi^i{}_k\Phi^k{}_{j}\chi^j\nn
&-\frac{1}{2}\xi_1\phi^2 R  -\xi_2\chi_i^{\dagger}\chi^i R.
\eea
Again $\Phi = T^a\phi^a$, where now $a= 1,2 \ldots N^2 - 1$. $\chi^i\
[i=1,2 \ldots N]$ is now a {\it complex\/} multiplet in the defining 
(fundamental)
representation, and $T^a$ are no longer (all) antisymmetric; 
they are again normalised so that
\beq
\Tr\, T^a T^b = \frac{1}{2}.
\eeq

\section{The $SU(N)$ $\beta$-functions}

\subsection{Classical gravity} 

As was the case with $SO(N),$ the potential \eqn{sunpot} still must include the 
$\xi_{1,2}$-terms. The flat space $\beta$-functions are~\cite{Einhorn:2017jbs}
\begin{subequations}\label{eq:betaflatB}
\begin{align}
\beta_{g^2}&=-b_g (g^2)^2,\ b_g\equiv \frac{21N{-}1}{3}
-\frac{4}{3}T_F,\\
\beta_{\lambda_1} &= (N^2{+}7)\lambda_1^2
{+}\frac{4(2N^2{-}3)}{N}\lambda_1\lambda_2
{+}\frac{12(N^2{+}3)}{N^2}\lambda_2^2\\
&+ 2N\lambda_4^2+4\lambda_4\lambda_5
-12Ng^2\lambda_1 {+} 18g^4,\\
\beta_{\lambda_2} &= \frac{4(N^2{-}9)}{N}\lambda_2^2 
+12\lambda_1\lambda_2+ \lambda_5^2
-12Ng^2 \lambda_2+3Ng^4,\\
\begin{split}
\beta_{\lambda_3} &= 2(N{+}4)\lambda_3^2+(N^2{-}1)\lambda_4^2
+\frac{(N{-}1)(N^2{+}2N{-}2)}{2N^2}\lambda_5^2\\
&+\frac{2(N^2{-}1)}{N}\lambda_4\lambda_5
- \frac{6(N^2{-}1)}{N} g^2 \lambda_3\\
&+ \frac{3(N{-}1)(N^2{+}2N{-}2)}{2N^2} g^4,
\end{split}\\
\beta_{\lambda_4} &= 4\lambda_4^2
+\lambda_4\bigg[(N^2{+}1)\lambda_1+\frac{2(2N^2{-}3)}{N}\lambda_2
+2(N{+}1)\lambda_3\bigg]\nonumber\\
&+
\lambda_5^2 + \lambda_5\bigg[\frac{N^2{-}1}{N}\lambda_1+
\frac{2(N^2{+}3)}{N^2}\lambda_2+2\lambda_3\bigg]\\
&-\frac{3(3N^2{-}1)}{N} g^2 \lambda_4 +3g^4,\nonumber\\
\begin{split}
\beta_{\lambda_5} &=\frac{N^2{-}4}{N}\lambda_5^2
+ \lambda_5\bigg[2\lambda_1+\frac{2(N^2{-}6)}{N}\lambda_2
+2\lambda_3+8\lambda_4\\
&-\frac{3(3N^2{-}1)}{N} g^2\bigg]
+3Ng^4.\end{split}
\label{sunbetas}
\end{align}
\end{subequations}
We also have 
\begin{subequations}\label{eq:betaxipA}
\begin{align}
\beta_{\xi'_1}^{(0)}&=\beta_{11}\xi'_1+\beta_{12}\xi'_2,\\
\beta_{\xi'_2}^{(0)}&=\beta_{21}\xi'_1+\beta_{22}\xi'_2,
\end{align}
\end{subequations}
where
\begin{subequations}\label{eq:betaij2B}
\begin{align}
\beta_{11}&=(N^2+1)\,\lambda_1{+}
2\frac{2N^2-3}{N}\lambda_2-6\alpha N,\\
\beta_{12}&=2N\lambda_4{+}\lambda_5,\\ 
\beta_{21}&=(N^2-1)\lambda_4+\frac{N^2-1}{2N}
\lambda_5,\\
\beta_{22}&=(2N+2)\lambda_3-3\frac{N^2-1}{N}\alpha.
\end{align}
\end{subequations}

\subsection{Gravitational Corrections}

As we indicated in the $SO(N)$ discussion above, the 
gravitational corrections to the matter $\beta$-functions have a
universal form; \eqn{eq:betamatter} remains valid 
in the $SU(N)$ case. 
There are minor changes to the gravitational corrections so that we have:
\begin{subequations}
\begin{align}
\label{eq:betagabcB}
\beta_a&=-b_2 a^2,& \beta_b&=-a^2 b_3(x,\xi'_1,\xi'_2),& \beta_c&{=}-b_1,\!\\
\label{eq:bgb1b2A}
b_2&= \frac{133}{10}+ N_a,&
b_3&\!\equiv\! \frac{10}{3}{-}5 x{+}\frac{5x^2}{12}{+}
\left((N^2-1)\xi'_1{}^2{+}
N\xi'_2{}^2\right)\!\frac{3x^2}{2},&
b_1&{=}\frac{196}{45}+ N_c.
\end{align}
\end{subequations}

\section{$SU(9)$ Fixed Points}

The $SU(N)$ discussion proceeds in the same manner as the $SO(N)$ case.  
In \reference{Einhorn:2017jbs} we showed that the minimum value of $N$ consistent with 
the existence of a UVFP is $N=9$, when $b_g^{\hbox{min}}$ is $4/3$. 
For the reduced couplings $x_i = \lambda_i/\alpha$ we found a flat space UVFP with 
\bea
x_1 &=& 0.386000,\ x_2 = 0.293121,\ x_3 = 0.502429,\nn
x_4 &=& 0.195158,\ x_5 = 0.398832.
\eea
Just as in the $SO(N)$ case, when gravitational corrections are included, we 
find a UVFP with similar $x_i$ values to the above, a large value of $x,$ and 
 {tiny nonzero values for $\xi'_1$ and $\xi'_2.$}

\section{$SU(N)$  Large $N$ limit}

We can define a large $N$ limit in a similar way to the $SO(N)$ case.
Once again we set
\beq\label{eq:rescalingflatA}
\ \alpha=\frac{\wt{\alpha}}{N},\ 
\lambda_1=\frac{\wt{\lambda}_1}{N^2},\  
\lambda_2=\frac{\wt{\lambda}_2}{N},\ 
\lambda_3=\frac{\wt{\lambda}_3}{N},\  
\lambda_4=\frac{\wt{\lambda}_4}{N^2},\ 
\lambda_5=\frac{\wt{\lambda}_5}{N}.
\eeq
As before there are distinct options for the rescaling of 
the gravitational corrections. 

\subsection{Gravitational coupling rescaling-A}

With identical rescalings to those described in the corresponding
$SO(N)$ case, we find  for the reduced quartic couplings $\wt{y}_i\equiv
\wt{\lambda}_i/\tilde{g}^2,$ :
\vskip-0.25in

\begin{subequations}\label{betabartildesuNflat2A}
\begin{align}
\betabar_{\wt{y}_1}&=\wt{y}_1^2+
12\wt{y}_2^2+18-(12-\wt{b}_g-8\wt{y}_2)\wt{y}_1,\label{SUN1}\\
\betabar_{\wt{y}_2}&=4\wt{y}_2^2
+3 -(12-\wt{b}_g)\wt{y}_ 2 ,\label{SUN2}\\
\betabar_{\wt{y}_3}&=2\wt{y}_3^2+
\frac{1}{2}\wt{y}_5^2 +
\frac{3}{2}-(6-\wt{b}_g)\,\wt{y}_ 3,\label{SUN3}\\
\begin{split}
\betabar_{\wt{y}_4}&=\wt{y}_5\big(\wt{y}_1+
2\wt{y}_2+2\wt{y}_3\big)+\wt{y}_5^2+3\\
&\hskip 0.1in +\wt{y}_4\big(
\wt{y}_1+4\wt{y}_2+2\wt{y}_3
-(9-\wt{b}_g)\big),\label{SUN4}
\end{split}\\
\betabar_{\wt{y}_5}&=
\wt{y}_5^2\!+
3-(9-\wt{b}_g-2\wt{y}_2)\wt{y}_ 5.\label{SUN5}
\end{align}
\end{subequations}
Just as in the $SO(N)$ case, any flat space  UVFP would  necessarily  imply  a corresponding FP 
with $\wt{\xi'_1} = \wt{\xi'_2} = x = 0$. We found such FPs for specific values of $\wt{b}_g$ in 
\reference{Einhorn:2017jbs}, as shown in Table~\ref{suUVFPsN}.

\begin{table}[hbt]
\begin{center}
\begin{tabular}{|c|c|c|c|c|c|} \hline
& & & & & \\
$\wt{b}_g$ & $\wt{y}_1$  & $\wt{y}_2$ & $\wt{y}_3$ & $\wt{y}_4$ & $\wt{y}_5$ \\ 
\hline
0.& 2.64270& 0.275255 & 0.289413  & 0.970346 &0.371374\\
\hline
$1/3$& 2.94605& 0.284989 & 0.312552 & 1.20422 &0.389234\\
\hline
1/2& 3.15683 & 0.290153 & 0.325788  & 1.39047 &0.398894\\
\hline
3/4&  3.67495& 0.298306 & 0.348280  & 1.94791 &0.414424\\
\hline
0.84798& 4.36728& 0.301646 & 0.358128  & 2.99190 &0.420885\\
\hline
\end{tabular}
\caption{\label{suUVFPsN} UVFPs for $SU(\infty).$}
\end{center}
\end{table}

Interestingly, \eqn{betabartildesuNflat2A}\ becomes {\it identical\/}  to \eqn{eq:betabarflatNA}, 
the corresponding set of results in the $SO(N)$ case, if we do the following:
\begin{itemize}
\item   Replace $\wt{y}_5$ by $\wt{y}_5/2$

\item   Replace $\wt{b}_g$ by $2\wt{b}_g$

\item    Multiply the RHS of all of \eqn{SUN1}-\eqn{SUN5} by 2.
\end{itemize}
At first sight, this result is surprising but can be understood (or at least made plausible),
by inspection of the respective Dynkin diagrams.  This means, of course that the FPs are 
essentially identical in the two cases. This is immediately apparent in the comparison of 
the first two rows of Table~\ref{soUVFPs}\ with the corresponding rows in Table~\ref{suUVFPsN}. 
(Recall that $\wt{b}_g= 1/6$ in the $SO(N)$ case corresponds to $\wt{b}_g= 1/3$  in the $SU(N)$ 
case).

In the same limit the gravitational couplings are:
\begin{subequations}\label{eq:betabarcurvedSUNA}
\begin{align}
\betabar_{\wt{a}} &= \wt{a}(\wt{b}_g -\wt{b}_2\wt{a}),\\
\betabar_{\wt{x}} &=  -\wt{a}\,\wt{x} \Big(\frac{3}{2}\wt{\xi'_1}^2\wt{x}
-\wt{b}_2\Big),\label {xNtildeSUNA}\\
\betabar_{\wt{\xi'_1}} &=(\wt{y}_1 +4\wt{y}_2-6)\wt{\xi'_1},\\
\betabar_{\wt{\xi'_2}} &=(2\wt{y}-3\-3)\wt{\xi'_2}
+(\wt{y}_4+\frac{1}{2}\wt{y}_5)\wt{\xi'_1}.
\end{align}
\end{subequations}
As we remarked above, any flat space  UVFP would  necessarily  imply  a corresponding FP 
with $\wt{\xi'_1} = \wt{\xi'_2} = x = 0$, but one destabilised by \eqn {xNtildeSUNA}.
However, once again we see from \eqn{xNtildeSUNA}  that  
such a FP is unstable with respect to fluctuations in $x$. 

Now let us  seek a FP for nonzero $\wt{\xi'}_1$. Just as in the $SO(N)$ case, we find two real FPs, as shown in 
Table~\ref{newsuFPs}.
\begin{table}[hbt]
\begin{center}
\begin{tabular}{|c|c|c|c|c|c|c|c|} 
\hline 
& & & & & & & \\
$\wt{b}_g$ & $\wt{y}_1$  & $\wt{y}_2$ & $\wt{y}_3$ & $\wt{y}_4$ 
& $\wt{y}_5$ & $x$ & $\wt{\xi'_2}$\\
\hline 
& & & & & & &  \\
0.8192 & 4.79737 &  0.30066 &  0.35517  & 3.90016 & 0.41896 &
${\wt{b}_2}/{(\wt{\xi'_1})^2}$ & $1.88636\wt{\xi'_1}$\\ 
\hline 
& & & & & & &  \\
0.40961 & 4.79737 &  0.30066& 2.23523&    -3.19276 & 0.41896 &
${\wt{b}_2}/{(\wt{\xi'_1})^2}$ & $1.88636\wt{\xi'_1}$\\ 
\hline 
\end{tabular} 
\caption{\label{newsuFPs} Curved Space UVFPs for $SU(\infty).$}
\end{center} 
\end{table} 
These results are essentially identical  to the corresponding ones for $SO(N)$; of course, 
the large $N$ $\beta$-functions in the two cases are essentially 
identical, differing only by redefinitions of the coupling $\wt{y}_5$  and overall rescaling. Once again 
we do not find any real FPs in this case. 

\subsection{Gravitational coupling rescaling-B}

We again have the possibility of a nontrivial  alternative rescaling, as
described for SO(N) in Section~\ref{subsec:NB}.  With \eqn{eq:rescalinggravB} 
we find (in the large $N$  limit):
\begin{subequations}\label{betabartildesuNflat2B}
\begin{align}
\betabar_{\wt{y}_1}&=\wt{y}_1^2+
12\wt{y}_2^2+18-(12-\wt{b}_g-8\wt{y}_2)\wt{y}_1+5\wt{a}^2\wt{\xi'_1}^2, \label{NB1}\\
\betabar_{\wt{y}_2}&=4\wt{y}_2^2
+3 -(12-\wt{b}_g)\wt{y}_ 2 ,\\
\betabar_{\wt{y}_3}&=2\wt{y}_3^2+
\frac{1}{2}\wt{y}_5^2 +
\frac{3}{2}-(6-\wt{b}_g)\,\wt{y}_ 3,\\
\betabar_{\wt{y}_4}&=\wt{y}_5\big(\wt{y}_1+
2\wt{y}_2+2\wt{y}_3\big)+\wt{y}_5^2+3\nn
&+ \wt{y}_4\big(
\wt{y}_1+4\wt{y}_2+2\wt{y}_3
-(9-\wt{b}_g)\big)
+5\wt{a}^2\wt{\xi'_1}\wt{\xi'_2},\\
\betabar_{\wt{y}_5}&=
\wt{y}_5^2\!+
3-(9-\wt{b}_g-2\wt{y}_2)\wt{y}_ 5,\nn
\betabar_{\wt{\xi'_1}} &= (\wt{y}_1 +4\wt{y}_2-6)\wt{\xi'_1}
+ \frac{10}{3}\frac{\wt{a}\wt{\xi'_1}}{\wt{x}}\\
\betabar_{\wt{\xi'_2}} &= \frac{1}{36}((-54+36\wt{y}_3)\wt{\xi'_2}
+18\wt{\xi'_1}(\wt{y}_4+\frac{1}{2}\wt{y}_5))
+ \frac{10}{3}\frac{\wt{a}\wt{\xi'_2}}{\wt{x}}\label{NB7}\\
\betabar_{\wt{x}} &=  -(1/12)\wt{a} (18\wt{x}^2\wt{\xi'_1}^2 
-12\wt{b}_2\wt{x}+40).\label{NB8}
\end{align}
\end{subequations}
Here, although the $\beta$-functions in \eqn{NB1}-\eqn{NB7}\ can be converted to the 
corresponding $SO(N)$ results by redefining $y_5$, $\wt{a}\xip_{1,2},$ and $x$, we cannot then 
redefine \eqn{NB8}\ in the same way. Nevertheless, we see that, in this case also, the natural 
generalisation of the UVFP we identified in flat space (corresponding to 
$\wt{\xi'_1} = \wt{\xi'_2} =0$) is destabilised by the gravitational corrections. 

\section{Dimensional Transmutation in the SO(N) model}\label{sec:dtson}

In either the $SO(N)$ or the $SU(N)$ case, a comprehensive analysis  of
the behaviour of the effective action with two distinct scalar
multiplets  and five independent quartic scalar couplings would be a formidable
undertaking. We  choose to make the crucial assumption that DT occurs
via the development of  a vev for the adjoint representation only, just as we
analysed in  \reference{Einhorn:2016mws}. The precise details differ,
however, because  the analysis involves the behaviour of the couplings
under renormalisation, and  the renormalisation of the adjoint self
couplings are, of course, affected by  the presence of the other
multiplet.  

In \reference{Einhorn:2016mws} we focused our attention on the $SO(10)$ theory.
 We showed it was asymptotically free both without and with gravitational  interactions,
and that in the latter case, the adjoint developed a vev via DT, with
symmetry breaking uniquely determined to be  $SO(10)\to SU(5)\otimes
U(1)$. With the addition of a multiplet of scalars in the 
fundamental representation, we showed in \reference{Einhorn:2017jbs}\ that 
asymptotic freedom is not sustained in the flat space case; the minimum value of $N$ necessary 
becomes $N=12$. We have seen above that this remains true when gravitational interactions are 
included. Therefore we need to generalise our previous discussion to accommodate $N>10$.

We begin by assuming that the background metric is well-approximated by the de~Sitter metric,
\beq
R_{\alpha\beta\gamma\delta} = \frac{R}{12}
\left(g_{\alpha\gamma}g_{\beta\delta}-g_{\alpha\delta}g_{\beta\gamma}\right),
\eeq
with constant $R>0.$ Then if $\Phi$ is constant and non-zero, we showed in 
\reference{Einhorn:2016mws}\ that, in the $SO(10)$ case, symmetry breaking 
occurs in the $SU(5)\otimes U(1)$ direction with 
\beq
\frac{\Phi}{\!\!\sqrt{NR}} =r\begin{pmatrix}
{\it 1}&\phantom{-} 0\\ 0 & -{\it 1}
\end{pmatrix},
\eeq
 where ${\it 1}$ is the $5\otimes5$ identity matrix, and 
 $r \equiv \sqrt{2T_2/(NR)}=\sqrt{(\phi^a\phi^a)/(NR)}.$

It is easy to see that if we assume that $\chi$ does not get a vev, then for 
any {\it even\/} $N$, $\Phi$ takes the same block form, that is proportional to 
\beq
r\begin{pmatrix}
{\it 1}& \phantom{-}0\\ 0 & -{\it 1},
\end{pmatrix}
\eeq
with $N/2\times N/2$ blocks.
The residual symmetry after such spontaneous symmetry breaking (SSB) is 
$SU(N/2)\otimes U(1)$ for \emph{arbitrary} $r.$ Thus the classical action for (even $N$)  
takes the off-shell value 
\beq
\label{eq:scloffshell}
\frac{S_{cl} (\lambda_i,r)}{V_4}=\frac{1}{3b}+\frac{c}{6}+\frac{\zeta_1}{2}(Nr^2)^2
- Nr^2\xi_1,
\eeq
where $V_4$ is an angular volume, and $\{\lambda_i\}$ is the complete set of couplings. Here, we introduce the symbol
$\zeta_1$ for the sum of the first two couplings,
\beq\label{eq:zeta1}
\zeta_1 \equiv \lambda_1 {+}2\lambda_2/N.
\eeq
In order for the action per unit volume to be bounded from below, we must have 
$\zeta_1>0.$  We shall also need the derivatives 
\beq\label{derivatives}
\frac{S'_{cl}}{V_4}=2 Nr\left[N\zeta_1 r^2-\xi_1 \right], \quad
\frac{S''_{cl}}{V_4}=2N\left[3N\zeta_1 r^2-\xi_1\right].
\eeq
Here (and subsequently)  
\beq\label{Sprime}
S'_{cl} \equiv \frac{\pa S_{cl} (\lambda_i,r)}{\pa r}.
\eeq

We can determine the extremal values of $r$ at tree level by solving 
$S'_{cl}=0.$ Therefore, $r=0$ or $r=r_0$ with 
\beq\label{eq:extremum}
r_0^2 \equiv \frac{\xi_1}{N\zeta_1},
\eeq
which, since $\zeta_1>0,$ yields $r_0$ real only if $\xi_1>0.$ Assuming this 
requirement is satisfied, the classical curvature $S''_{cl}$ is negative at $r=0,$ 
so the unbroken solution is a maximum.  At $r_0,$ the curvature takes the value 
\beq \label{eq:sclonshellB}
\frac{S_{cl}^{''(os)}}{V_4}= 4N\xi_1>0.
\eeq
At the minimum, the value of the classical action is
\beq \label{eq:sclonshellA}
\frac{S_{cl}^{(os)}}{V_4}=\frac{1}{3b}+\frac{c}{6}
-\frac{\xi_1^2}{2\zeta_1}.
\eeq

Recall that, for $N=12,$ the UVFP, \eqn{uvfpgrav}, has $\xi'_1\approx 0,$ or 
$\xi_1\approx -1/6.$ This is a generic result, so it is important to establish that 
it is possible to fulfill this condition and still have $\xi_1 > 0$ at the DT scale. 
(This did in fact occur the simpler model~\cite{Einhorn:2016mws}
that did not include the scalar $\chi$ in the fundamental.)

Generally, we want to determine whether DT can occur. Including
radiative corrections,  the effective action takes the generic form 
\beq \label{eq:effaction} 
\Gamma(\lambda_i,r,\rho/\mu) =
S_{cl}(\lambda_i,r) + B(\lambda_i,r)\log(\rho/\mu) +
\frac{C(\lambda_i,r)}{2}\log^2(\rho/\mu) +\ldots, 
\eeq 
where $\rho\equiv \sqrt{R}.$ All coupling constants are denoted by the set
$\{\lambda_i\}.$ In writing the effective action in this form, we have
assumed that  $\Phi$ is spacetime independent; 
it is not necessary to assume that the breaking pattern is to
$SU(N/2)\otimes U(1).$ 

We seek an extremum of the effective action such that 
$(r,\rho)= (\vev{r},\vev{\rho})\equiv (r_0,v)$, i.e., solutions to
\begin{subequations} \label{eq:eomonshell} \begin{align}
\label{eq:eomronshell} \frac{\partial}{\partial r}\Gamma(\lambda_i, r,
\rho/ \mu)\Big|_{r_0,v}&= \frac{\partial}{\partial
r}S_{cl}(\lambda_i,r)\Big|_{r_0,v}=0,\\ \label{eq:eomrhoonshell}
\rho\frac{\partial}{\partial\rho} \Gamma(\lambda_i, r,
\rho/\mu)\Big|_{r_0,v}&=B(\lambda_i,r)\Big|_{r_0,v}=0, \end{align}
where we choose $\mu = \vev{\rho} \equiv v$. 
\end{subequations}
These results are exact to all orders in the loop expansion.

In order to determine stability, we shall also need the matrix of second derivatives on-shell:
\begin{subequations}
\label{eq:secondonshell}
\begin{align}
\frac{\partial^2}{\partial r^2}\Gamma(\lambda_i, r, \rho/ \mu)\Big|_{r_0,v}&=
\frac{\partial^2}{\partial r^2}S_{cl}(\lambda_i,r)\Big|_{r_0,v},\\
\rho\frac{\partial^2}{\partial r\partial\rho}\Gamma(\lambda_i, r, \rho/ \mu)\Big|_{r_0,v}&=
\frac{\partial}{\partial r}B(\lambda_i,r)\Big|_{r_0},\\
\rho^2\frac{\partial^2}{\partial \rho^2}\Gamma(\lambda_i, r, \rho/ \mu)\Big|_{r_0,v}&=
C(\lambda_i,r_0).
\end{align}
\end{subequations}
As before, we implicitly set $\mu=v$ after performing the derivatives.
The second variation on-shell can be written 
\beq\label{eq:deta2onshell}
\delta^{(2)}\Gamma=\frac{1}{2}
\begin{pmatrix}
\frac{\delta\rho}{\rho} & \delta r
\end{pmatrix}
\begin{bmatrix}
C(\lambda_i,\!r_0) &\ B{}^\prime(\lambda_i,\!r_0)\\
B{}^\prime(\lambda_i,\!r_0) &\ S_m^{''}(\lambda_i,\!r_0)
\end{bmatrix}
\begin{pmatrix}\frac{\delta\rho}{\rho} \\ \delta r
\end{pmatrix}.
\eeq
Given our conventions, these equations \eqn{eq:secondonshell} are also 
exact to all orders in the loop expansion, but their leading nonzero 
contributions vary from tree level for those involving $S_{cl}$, 
to one-loop for $B$, to two-loop\footnote{However,  the two-loop contribution 
to $C_2$ can be calculated from one-loop corrections; to $C_3$, 
from two-loop corrections, etc.} for $C$. This is the characteristic 
``see-saw" pattern, so this matrix has two eigenvalues $\varpi_i$ that may be 
approximated as
\beq\label{eq:eigenvalues}
\varpi_1(r_0,v)=\frac{ S_{cl}^{\,''}}{2}+O(\hbar^2),\qquad 
\varpi_2(r_0,v)=\half\left[C_2
-\frac{\left(B'_1\right)^2}{S_{cl}^{''}}\right]+O(\hbar^3).
\eeq
$\varpi_1$ is determined by the classical curvature and given by
\eqn{eq:sclonshellB}, and $\varpi_2$, although of order 
$\hbar^2$, is determined by one-loop results.

In \reference{Einhorn:2014gfa}, we used the renormalisation group to show 
that (to leading order)
\beq\label{eq:b1os}
B_1=\sum_i\beta_{\lambda_i}\frac{\pa S_{cl}}{\pa\lambda_i}
\eeq
so that  the conditions  for an extremum
corresponding to DT in this model (and others of this general form) are
$r=r_0$ and 
\beq\label{eq:b1osa}
B_1\Big|_{r=r_0}= B_1^{(os)} = \sum_i\beta_{\lambda_i}\frac{\pa S_{cl}}
{\pa\lambda_i}\bigg|_{r=r_0}
=\sum_i\beta_{\lambda_i}\frac{\pa S_{cl}^{(os)}}{\pa\lambda_i}=0
\eeq
where $B_1^{(os)}, S_{cl}^{(os)}$
 are  the ``on-shell'' contributios to $B_1$, $S_{cl}$ with
``on-shell'' corresponding to $r=r_0$.
Notice that 
\beq
\frac{\pa S_{cl}^{(os)}}{\pa\lambda_i}=
\left[\frac{\pa S_{cl}}{\pa\lambda_i} + 
\frac{\pa S_{cl}}{\pa r} \frac{\pa r}{\pa \lambda_i}\right]\bigg|_{r=r_0}= 
\frac{\pa S_{cl}}{\pa\lambda_i}\bigg|_{r=r_0}.
\eeq
Such an extremum corresponds
to a minimum if $\zeta_1 > 0$ and $\varpi_2 > 0$, 
and  we showed, again using the renormalisation group, 
that~\cite{Einhorn:2014gfa,Einhorn:2015lzy,Einhorn:2016mws}
\beq\label{eq:varpi2A}
\varpi_2=\half\left[\beta_{\lambda_i}^{(1)}\frac{\partial B_1}{\partial \lambda_i}
-\frac{1}{S_{cl}^{''}}\left(\!\beta_{\lambda_i}^{(1)}
\frac{\partial}{\partial \lambda_i } S'_{cl}(\lambda_i,r)
\!\right)^{\!2}  \right]\Bigg|_{r_0,v}.
\eeq
or using \eqn{eq:b1os},
\beq\label{eq:varpi2B}
\varpi_2=\half\left[\left(\beta_{\lambda_i}^{(1)}
\frac{\partial}{\partial \lambda_i }\right)^{\!2}\! S_{cl}
-\frac{1}{S_{cl}^{''}}\left(\!\beta_{\lambda_i}^{(1)}
\frac{\partial}{\partial \lambda_i } S'_{cl}(\lambda_i,r)
\!\right)^{\!2}  \right]\Bigg|_{r_0,v}.
\eeq
Remarkably, $\varpi_2$ can also be written
\beq\label{eq:varpi2C}
\varpi_2=\half\left[\beta_{\lambda_i}^{(1)}
\frac{\partial B_1^{os}}{\partial \lambda_i}\right],
\eeq
an observation which is not particularly obvious and that we shall explain in 
detail elsewhere in a more general discussion of the RG and effective actions 
in this kind of theory.

From \eqn{eq:b1osa}, we find
\beq\label{b1off2}
B_1 (\lambda_i,r)=\frac{b_3(x,\xi'_1,\xi'_2)}{3x^2}-\frac{b_1}{6}
+\half \big(\alpha N r^2\big)^2 \big(\betabar_{z_1}-b_g z_1\big)
 -(\alpha N r^2)\betabar_{\xi'_1},
\eeq 
where $b_1, b_3$ are given in \eqn{eq:b1b2b3}, and $z_1=\zeta_1/\alpha,$ 
where $\zeta_1$ was defined in \eqn{eq:zeta1}. (Obviously, 
$\overline{\beta}_{\xi'_1}=\overline{\beta}_{\xi_1}.$)
\beq\label{eq:b1osN}
B_1^{(os)}=\frac{b_3(x,\xi'_1,\xi'_2)}{3x^2}-\frac{b_1}{6}+
\Big(\frac{\xi_1}{z_1}\Big) \left[\half\Big(\frac{\xi_1}{z_1}\Big) 
\big(\betabar_{z_1}-b_g z_1\big) -\betabar_{\xi'_1}\right].
\eeq
This rather succinct formula hides a good deal of complexity in the 
expressions for $\betabar_{z_1} = \betabar_{x_1} + 2\betabar_{x_2}/N$ (see 
\eqn{eq:betabars1c} for the $N=12$ case),  and for 
$ \betabar_{\xi'_1},$ \eqns{betabarxipk}{deltabetabarxipk}. We shall leave 
further discussion of the determination of the extrema from $B_1^{(os)}=0$ 
and the flow from such an extremum towards the UVFP to the next section, 
\secn{sec:dtso12}.

\section{Dimensional Transmutation in the SO(12) model}\label{sec:dtso12}

Even after restriction to $N=12, b_g=1/6,$ this relatively simple model is still extremely complicated, 
both analytically and numerically, for several reasons.  In our classically scale-invariant version, it 
involves many fields, and only the possible ``directions" of SSB in this field 
space is determined, depending on relations among the 10 dimensionless coupling constants of the 
bosonic sector. The scale of SSB is determined at one-loop order, and the determination of 
its character (and correspondingly, the dilaton mass) is determined at two-loops. The numerical results
for the UV behavior were given in \secn{sec:so12}, showing that such a model can be AF for a certain 
range of coupling constants.

We now wish to show that the model can undergo DT and that it is locally stable for at least a portion of 
the DT surface. This will require fleshing out \eqn{eq:b1osN}  in greater detail. From \eqn{eq:b1b2b3}, we have for $N=12,$
\beq\label{eq:b3}
b_3=\frac{10}{3}{-}5 x{+}\frac{5x^2}{12}{+}
9\left(11\xi'_1{}^2{+}
2\xi'_2{}^2\right)x^2,
\eeq
As mentioned 
in \secn{sec:so12}, even specifying $b_g=1/6$ does not yield a unique model, since several different 
arrangements of the fermion content are still possible. In \secn{sec:so12}, we arbitrarily chose to focus 
on the case of $52$ two-component fermions in the fundamental representation $N$ so that $N_f=312$ 
and $b_1=917/36, b_2=59.$ As noted earlier, as the running scale decreases, we must hope that 
$\xi'_1$ will run from near zero to the region where $\xi_1=\xi'_1-1/6>0.$ For this reason, exploring the 
DT surface may be simpler in terms of $\xi_1$ rather than $\xi'_1.$
Further, $B_1$ depends on the couplings $(x_4,x_5)$ only via the linear combination\footnote{For general $N$, $z_4\equiv x_4+x_5/(2N)$).}
$x_4+x_5/24\equiv z_4.$   After making these notational changes, we find that $B_1^{(os)},$ 
\eqn{eq:b1osN}, becomes
\begin{align}\label{eq:b1os12t}
\begin{split}
B_1^{(os)}&= \frac{10}{9x^2}-\frac{5}{3x}-\frac{689}{216}+\frac{\xi_1}{3}(6\xi_1{-}1)
+6\xi_2^{'2}
+\frac{\xi_1}{z_1}\bigg[5{-}\frac{35 x_2}{18}{-}12\xi'_2 z_4\bigg]\\
&~~~+\frac{\xi_1^2}{z_1^2}\bigg[5+\frac{35 x_2^2}{18}+6 z_4^2 \bigg]
+\frac{\abar\, \xi_1^2}{2z_1}\bigg[5+\frac{x}{6}-\frac{20}{3x}-x\xi_1(12 \xi_1{+}1)\bigg]\\
&~~~+\frac{\abar^2 \xi_1^4}{2z_1^2}\bigg[5+\frac{x^2}{4}+3x^2\xi_1(3\xi_1{+}1)\bigg].
\end{split}
\end{align}
We note that $z_1$ enters only as the ratio $\xi_1/z_1,$ and $\abar$ appears only as the 
combination $\abar\xi_1^2/z_1,$ which may well turn out to be very small.

To determine $\varpi_2,$ we need to evaluate either \eqns{eq:varpi2A}{eq:varpi2B} or 
\eqn{eq:varpi2C} for $N{=}12.$ 
From \eqn{eq:b1os12t}, we see that, when expressed in terms of the 7 rescaled variables 
$\{\abar, x, \xi_1,\xi'_2, z_1,x_2,z_4\},$ $B_1^{(os)}$ is independent of 
$\{x_3, x_5\}.$ Consequently, we require 7 reduced $\beta$-functions. The actual 
determination of $\varpi_2$ is unavoidably complicated and not very illuminating. 
For completeness, In Appendix~\ref{sec:varpi2}, we give the relevant $\beta$-functions
and the resulting exact formula for $\varpi_2.$ 

For our purpose here, we shall be satisfied to demonstrate 
a region of parameter space such that $\varpi_2 > 0$ {\it and\/} our other requirements 
(such as $\xi_1 > 0$) are satisfied {\it and\/} that 
the couplings there proceed to run to the UVFP. (A complete 
description of solutions of $B_1^{(os)}=0$  remains a formidable undertaking).
The simplest way to identify such a region is to start with as many couplings as possible 
(consistent with requirements at the DT scale)  already near the UVFP. The most obvious 
problem with this is that, since we plan to generate the E-H term using the adjoint vev, 
we require $\xi_1>0$, whereas at the UVFP $\xi_1 \approx -1/6$. We can however 
choose to try $\xi'_2$ at or near zero.

In Table~\ref{DTsurface}, we give two examples of points satisfying 
$B_1 = 0$, $\varpi_2 > 0,$ and flowing to the UVFP.  
\begin{table}[hbt]
\begin{center}
\begin{tabular}{|c|c|c|c|c|c|c|c|} 
\hline 
& & & & & & & \\
$\wt{x}_1$  & $\wt{x}_2$ & $\wt{x}_3$ & $\wt{x}_4$ 
& $\wt{x}_5$ & $x$ & $\wt{\xi_1}$ & $\wt{\xi'_2}$\\
 & & & & & & & \\\hline
& & & & & & &  \\
0.373073 & 0.25 &  0.377518 &  0.25 & 0.582159 & 148.271 &
0.190359 & 0.042969\\ 
\hline 
& & & & & & &  \\
0.3577213 &   0.25& 0.377518 &    0.25 & 0.582159& 120.471&
0.184093 & 0\\ 
\hline 
\end{tabular} 
\caption{\label{DTsurface} DT points in the UVFP catchment basin.}
\end{center} 
\end{table}

\section{Summary and Conclusions}

This paper continues a series in which we have developed the theory of renormalisable 
quantum gravity (RQG) coupled to matter fields, including in particular a clear demonstration 
that generation of scalar vacuum expectation values can occur via Dimensional 
Transmutation (DT). We have also claimed to identify a large class of such theories that 
exhibit asymptotic freedom, and consequently represent a possible UV completion of Einstein 
gravity (in combination with DT).  This claim remains controversial because 
of disagreement~\cite{Salvio:2014soa}\ concerning the correct sign of the coefficient 
$b$ of the $R^2$ term in the Lagrangian, \eqn{eq:sgrav}.
We believe that the sign we adopt (which results in AF for the $b$-coupling) is required 
for convergence of the Path Integral.

In another  previous paper~\cite{Einhorn:2017jbs}, we addressed the issue of AF in 
$SU(N)$ and $SO(N)$ gauge theories in flat space in the presence of adjoint and 
fundamental scalar representations. We identified some errors in the original treatment, and 
also added a discussion of a large $N$ limit, where for a limited range of  (small) values of 
$\tilde{b}_g$ we showed that a UVFP existed. 

In this paper we generalise this work by coupling these theories to RQG. In both 
the $SO(N)$ and $SU(N)$ cases, the inclusion of these interactions has essentially no effect 
on the minimum value of $N$ required for a UVFP, nor in the basic features of this FP. We 
described the $SO(10)$  case (with only an adjoint scalar)  in detail in 
\reference{Einhorn:2016mws}.  With 
both scalar representations present, however, the minimum value of $N$ for AF becomes 
$N=12$.\footnote{ In the case of $SU(N),$ the minimum value of $N$ for a UVFP 
is $N=7$ for the theory with only the adjoint scalar, becoming $N=9$ in the presence of 
both adjoint and fundamental.} We argued, that in the $SO(12)$ case, the theory with 
both scalar representations would 
exhibit a region of parameter space consistent with Dimensional Transmutation, in a 
similar way to the case with an adjoint only, and, moreover, that from parts of this region, 
the couplings flowed to a UVFP (which we identified) at high energies. In 
Appendix~\ref{sec:varpi2}, we present the expression for $\varpi_2$, defined in 
Eqs.~(\ref{eq:varpi2A})-(\ref{eq:varpi2C}), which is required to be positive within
a subregion of the DT surface.

The large $N$ limit is interesting. We identified two distinct ways to implement the limit for the 
gravitational couplings. In both cases, however, the UVFP that we identified in flat space is 
destabilized by inclusion of these couplings.
 
To sum up: we have demonstrated that it is feasible to construct a realistic Grand Unified Theory 
with a non-trivial set of scalar field representations (adjoint+scalar) for both $SU(N)$ and $SO(N)$ cases, 
with complete asymptotic freedom and (we argue) Dimensional Transmutation to a low energy theory 
with gravitational self-interactions  described by the Einstein term. However the cases $SO(10)$ 
and $SU(5)$ are excluded. Moreover, the minimum value of $N$ required is higher
for both $SU(N)$ and 
$SO(N)$ than in the case with only an adjoint scalar, making it not unlikely that it will be 
higher still with a more complicated set of scalar representations. Problems with the scenario 
remain, most obviously the generation of the electroweak scale, and the issue of unitarity, which 
we hope to address elsewhere~\cite{EJ2019}. 

\begin{acknowledgments}
DRTJ thanks  KITP (Santa Barbara), where part of this work was done, 
for hospitality, and the Leverhulme Trust for the award of an 
Emeritus Fellowship. This research was supported in part
by the National Science Foundation under Grant 
No.\ NSF \nolinebreak PHY11-25915, by the Leverhulme Trust, and by the University of Liverpool. 
\end{acknowledgments}

\begin{appendix}

\section{Reduced $\beta$-functions and result for $\varpi_2$}\label{sec:varpi2}

With reference to \secn{sec:dtso12}, the relevant $\beta$-functions 
for $N{=}12, b_g{=}1/6$ are
\begin{subequations}
\begin{align}
\betabar_\abar\,&=59\abar\Big(\frac{1}{354}-\abar\Big),\\
\betabar_x\,&=\abar\Big(\!-\frac{10}{3}+64x -x^2\Big(\frac{5}{12}+
9\big(11\Big(\xi_1{+}\frac{1}{6}\Big)^2{+}2\xi'_2{}^2\big)\Big)\Big), \\
\betabar_{z_1}&=\abar^2\xi_1^2\Big(5+9x^2\Big(\xi_1{+}\frac{1}{6}\Big)^2\,\Big)+
\abar z_1\Big(5{-}18x \Big(\xi_1{+}\frac{1}{6}\Big)^2\,\Big)+74 z_1^2\nn 
& +\frac{z_1}{6}(140x_2{-}359)+\frac{35 x_2^2}{9}+12z_4^2+10,\label{eq:betabars1c}\\
\betabar_{x_2}&=\abar x_2\Big(5{-}18x \Big(\xi_1{+}\frac{1}{6}\Big)^2\,\Big)+21 x_2^2+
x_2\Big(12z_1{-}\frac{359}{6}\Big)+\frac{x_5^2}{8}+6,\\
\betabar_{z_4}&=\abar^2\xi_1\Big(\xi'_2{-}\frac{1}{6}\Big)
(5{+}9x^2\Big(\xi_1{+}\frac{1}{6}\Big)\xi'_2)+\abar z_4\big(5{-}3x\Big(\Big(\xi_1
{+}\frac{1}{6}\Big)^2{+}4\Big(\xi_1{+}\frac{1}{6}\Big)\xi'_2{+}\xi'_2{}^2\Big)\big)\nn
&+4z_4^2+z_4\Big(68z_1{+}14x_3{+}\frac{35 x_2}{3}{-}\frac{139}{3}\Big)
+\frac{35 x_5^2}{144}+\frac{5}{2}, \\
\betabar_{\xi_1}&=\abar \xi_1 
\Big(\frac{10}{3x} {-}  x \Big(\xi_1{+}\frac{1}{6}\Big) (3\xi_1{+}2)\Big)
+ \Big(\xi_1{+}\frac{1}{6}\Big) \Big(68 z_1{+}\frac{35 x_2}{3}{-}30 \Big)+ 12 \xi'_2 z_4,  \\
\betabar_{\xi'_2}&= \abar\Big(\xi'_2{-}\frac{1}{6}\Big)\Big(\frac{10}{3 x} {-} 
\frac{3}{2} x \xi'_2\big(1 {+} 2 
\xi'_2\big)\Big)+\xi'_2 \Big(14 x_3-\frac{33}{2}\Big)+12 \Big(\xi_1{+}\frac{1}{6}\Big) z_4.
\end{align}
\end{subequations}
Since $x_3$ and $x_5$ do not appear in $B_1^{(os)},$ \eqn{eq:b1os12t}, 
we have omitted $\betabar_{x_3}$ and $\betabar_{x_5}$ above. Here, as with 
$B_1^{(os)},$ we have employed an asymmetric notation in terms of $\{\xi_1,\xi'_2\}.$

$\varpi_2$ is $\alpha$ times a function of all 9 rescaled couplings. (Recall our 
convention is $\alpha\equiv g^2.$) After restoring factors of $16\pi^2,$ the result of the 
evaluation of \eqn{eq:varpi2C} is\footnote{For readers who would
prefer this formula in Mathematica or Maple format, we have provided these 
online as supplementary files.} 
\begin{align}
&\varpi_2=\frac{\alpha}{512\pi^4}\bigg[\frac{5}{3}(7 x_2{-}18)\!\left(\! 4 \xi_1^2{-}
\frac{35}{18}\right)+6 \xi'_2{}^2 (28 x_3{-}33)- 
\abar^4 \frac{\xi_1^6}{z_1^3} \left(5 + \frac{x^2 (6 \xi_1{+}1)^2}{4}\right)^2\!\!+\nn &
\abar^3 \frac{\xi_1^4}{z_1^2}
\left(\frac{50 }{x}-\frac{5}{8} x^2  (6\xi_1{+}1)^2+\frac{5}{12} x  (6\xi_1{+}1)(72\xi_1+1)-
\frac{665 }{2}-\right. \nn &
\left.\frac{1}{48} x^3 \xi_1^4 (6\xi_1{+}1)^2 \left(41+402 \xi_1+
1008 \xi_1^2+216 \xi'_2{}^2\right)\right)+\nn &
{\frac{34}{9} (6\xi_1{+}1)(12 \xi_1{-}1) z_1-8 \xi'_2 z_4+\xi_1 \left(96 \xi'_2 z_4-40-
\frac{70 x_2}{9}\right)-}\nn &
\frac{\xi_1^2}{81 z_1^3} \left(108 z_4^2+35 x_2^2+90\right)^2{+}
\frac{\xi_1}{648 z_1^2} 
\Big(8 (35 x_2+216 \xi'_2 z_4-90) \left(108 z_4^2+35 x_2^2+90\right)+\nn &
3 \xi_1\big(7840 x_2^3{-}25200 x_2^2{+}105 x_2 \big(x_5^2{-}192\big){+}18 
\big(5 z_4 \big(7 x_5^2{+}72\big){+}72 (28 x_3{-}33) z_4^2+
576 z_4^3{+}3580\big)\big)\Big){+} \nn &
\abar^2 
\left(\frac{\xi_1^2}{z_1}\left(\frac{1330 }{3 x}-\frac{100 }{3 x^2}+\frac{5x}{12} 
(6\xi_1{+}1)(12\xi_1{-}1) -
\frac{5}{18}  \left(611+12 \xi_1+432 \xi_1^2-36 \xi'_2+432 \xi'_2{}^2\right)+
\right.\right. \nn &
\left. \frac{x^2}{18}   (6\xi_1{+}1) 
\left(1719 \xi_1^2+5184 \xi_1^3+27 \xi'_2{}^2-324 \xi'_2{}^3+9 \xi_1 
\left(13+36 \xi'_2{}^2\right)-5   \right)\right)-\nn &
{\frac{\xi_1^4}{z_1^3}\left(\frac{10}{9}  \left(108 z_4^2+35 x_2^2+90\right){+}\frac{x^2}{18}   (6\xi_1{+}1)^2
\left(108 z_4^2+35 x_2^2+90\right)\right)+}\nn &
\frac{\xi_1^3}{z_1^2}\left(\frac{5}{6}\big(35 x_2{+}12z_4(24 \xi'_2{-}1) {-}90\big){+}
\frac{ x^2}{24}  (6\xi_1{+}1) 
\left(35 x_2{+}60 \xi_1^2 (7 x_2{-}18){+}8 \xi_1 (35 x_2{+}216 \xi'_2 z_4{-}90){+}
\right.\right.\nn &
\left.\left.\left.18 \left(8 \xi'_2 z_4(3 \xi'_2{+}1)-5\right)\right)\right)\right){+}
{\frac{1}{z_1}\left(\xi_1^2 
\left(\frac{70 }{3}x_2^2-48 z_4^2-60\right)-
\frac{1}{324} (35 x_2{+}216 \xi'_2 z_4{-}90)^2+\right.}\nn &
{\left.\xi_1 \left(\frac{1505}{6}+\frac{700 x_2^2}{27}-\frac{35 x_5^2}{144}+
4 z_4^2-\frac{1}{12} \xi'_2 \left(35
x_5^2+72 \left(5-66 z_4+56 x_3 z_4+8 z_4^2\right)\right)\right)\right)+}\nn &
\abar \left(\frac{200}{27 x^3}-\frac{1330}{9 x^2}-\frac{5}{18} \left(19-6 \xi_1
+36 \xi_1^2+108 \xi'_2{}^2\right)+\right.
{\frac{10}{27 x} \left(307-9 \xi_1+72 \xi_1^2-18 \xi'_2+216 \xi'_2{}^2\right)+}\nn &
x \left(2 \xi_1-\frac{37 \xi_1^2}{3}-346 \xi_1^3-1200 \xi_1^4-3 \xi'_2{}^2 
(6\xi'_2{-}1)(2\xi'_2{+}1)\right)+\nn & 
\frac{1}{z_1}
\bigg(\frac{5}{9 x}\xi_1\left(90-35 x_2+12(1-24 \xi'_2) z_4\right){+}
\frac{5}{18} \xi_1 (35 x_2+36(6 \xi'_2 z_4{-}5))+ 
x \bigg(\!-20 \xi_1^4 (7 x_2{-}18) - \nn &
\frac{2}{3} \xi_1^3 (70 x_2{+}648 \xi'_2 z_4{-}315){+} 
\xi_1^2 \big(25{+}\frac{35}{18} x_2{+}48 \xi'_2 z_4(3\xi'_2{-}1)\big){+} 
\xi_1 \Big(\frac{35}{36} x_2{+}2 \xi'_2 z_4\big(18 \xi'_2{+}36 \xi'_2{}^2{-}1\big) 
\!\Big)\!\bigg){+}\!\nn &
\frac{\xi_1^2}{z_1^2}\bigg(\frac{20\left(108 z_4^2+35 x_2^2
+90\right)}{27 x}{-}\frac{5}{18}\left(35 x_2^2+54 \big(2 z_4^2{+}5\big)\right){+}
\frac{x}{108}\Big(\!90-175 x_2^2-
1296 z_4^2\xi'_2(3\xi'_2{+}2)+  \nn & 
36 \xi_1^2 \left(630+35 x_2^2+648 z_4^2\right)-
24 \xi_1 \big(35 x_2^2+ 18 (9 (4 \xi'_2-1)z_4^2-10)
\big)\Big)\bigg) \bigg)\bigg].
\end{align}

\end{appendix}

\end{document}